\documentclass[twocolumn,showpacs,preprintnumbers,amsmath,amssymb,floatfix]{revtex4}
\usepackage{epsf}
 \newcommand{\insertplot}[5]{\begin{figure}
 \hfill\hbox to 0.05in{\vbox to #5in{\vfill
 \inputplot{#1}{#4}{#5}}\hfill}
 \hfill\vspace{-.1in}
 \caption{#2}\label{#3}
 \end{figure}}
 \newcommand{\inputplot}[3]{
 \special{ps: plotfile #1}
}
\begin{document}
  \title{Horizon Properties of Einstein-Yang-Mills Black Holes}
\author{
{\bf Burkhard Kleihaus}
}
\affiliation{
{Department of Mathematical Physics, University College, Dublin,
Belfield, Dublin 4, Ireland}
}
\author{
{\bf Jutta Kunz}
}
\author{
{\bf Abha Sood}
}
\altaffiliation{
{GKSS, D-21502 Geesthacht, Germany}
} 
\author{
{\bf Marion Wirschins}
}
\affiliation{
{\small Fachbereich Physik, Universit\"at Oldenburg, 
D-26111 Oldenburg, Germany}
}
\date{\today}
%
%
%
\begin{abstract}
We consider static axially symmetric 
Einstein-Yang-Mills black holes in the isolated horizon formalism.
The mass of these hairy black holes is related
to the mass of the corresponding particle-like solutions 
by the horizon mass.
The hairy black holes violate the ``quasi-local uniqueness conjecture'',
based on the horizon charges.
\end{abstract}
\maketitle
 \vfill
 \vfill\eject

\section{Introduction}

In Einstein-Maxwell (EM) theory black holes are completely characterized 
by their mass, their electric and magnetic charge and their angular momentum,
i.e.~EM black holes have ``no hair''
\cite{hair-em}. 
The unique family of stationary Kerr-Newman black holes includes,
beside the stationary charged black holes,
the stationary Kerr black holes for vanishing electromagnetic charges,
the static Reissner-Nordstr\o m (RN) black holes 
for vanishing angular momentum,
and the static Schwarzschild black holes,
which carry neither electromagnetic charge nor angular momentum.
Notably, the stationary EM black hole solutions are axially symmetric,
while the static EM black hole solutions are spherically symmetric
\cite{hair-em}.

The uniqueness theorem (``no hair'' theorem) for black holes in EM theory
does not generalize to theories with non-abelian fields coupled to gravity.
Non-abelian black holes
are not completely determined by their global charges,
defined at infinity \cite{su2,review}.
Furthermore, 
non-abelian static black holes need not be spherically symmetric,
but might possess axial symmetry \cite{kk2,map3,hkk}
or discrete symmetries \cite{ewein} only,
i.e.~Israel's theorem does not generalize to
theories with non-abelian fields coupled to gravity either.

In particular, SU(2) Einstein-Yang-Mills (EYM) theory
possesses besides embedded abelian black holes
sequences of genuinely non-abelian black holes \cite{su2,kk2,rot}.
These hairy black holes
possess non-trivial matter fields outside their regular event horizon.
Their global charges alone do not determine the spacetimes uniquely.
Instead, they are additionally characterized by two integers,
the winding number $n$ (with respect to the azimuthal angle $\phi$)
of the non-abelian gauge fields
and the node number $k$ of the gauge field functions
\cite{kk2}.

Beside black hole solutions EYM theory also possesses
globally regular solutions \cite{bm,review,kk1}.
These particle-like solutions are obtained from the hairy
black hole solutions in the limit of vanishing horizon size.
Therefore the particle-like solutions are also characterized
by the winding number $n$ and the node number $k$
of the gauge fields.

Recently considerable progress concerning the understanding of
the non-abelian black hole solutions has been made in the
framework of the newly developed isolated horizon formalism
\cite{iso,iso1,iso2}.

In the isolated horizon formalism one considers space-times
with an interior boundary, which satisfy quasi-local boundary conditions
that insure that the horizon remains isolated \cite{iso,iso1,iso2}.
The boundary conditions imply that quasi-local charges can be defined
at the horizon, which remain constant in time.
In particular one can define a horizon mass, a horizon electric
charge and a horizon magnetic charge \cite{iso1}.

Amazingly, the horizon mass of hairy black holes 
is related in a simple way to their ADM mass
and to the ADM mass of the corresponding particle-like solution 
\cite{iso,iso1,iso2}.
This observation suggests the interpretation of hairy black holes
as bound states of a particle-like solution and a Schwarzschild black hole
\cite{iso2}.
Most interestingly, a ``quasi-local uniqueness conjecture''
for black holes has been proposed, stating
that static black holes are uniquely determined
by their horizon area and their horizon electric and magnetic charges
\cite{iso1}.

Here we address these issues, raised from the perspective of the
isolated horizon framework, for the static SU(2) EYM solutions.
We focus on static axially symmetric black hole solutions,
since static spherically symmetric black hole
solutions have been considered before \cite{iso,iso1,iso2}.
We verify the mass formula for these static axially symmetric black holes,
and consider the bound state interpretation.
Our main concern is, however, the ``quasi-local uniqueness conjecture'',
since, as a first check for this conjecture,
Corichi, Nucamendi and Sudarsky \cite{iso1} have proposed to study
colored black holes which are static but not spherically symmetric.

\section{\bf Static axially symmetric solutions}

We consider the SU(2) EYM action
\begin{equation}
S=\int \left ( \frac{R}{16\pi G} 
 - \frac{1}{2}{\rm Tr} (F_{\mu\nu} F^{\mu\nu})
 \right ) \sqrt{-g} d^4x
\   \end{equation}
with
$F_{\mu \nu} = 
\partial_\mu A_\nu +\partial_\nu A_\mu - i e \left[A_\mu , A_\nu \right] $,
Yang-Mills coupling constant $e$, and Newton's constant $G$.

The static axially symmetric solutions are obtained in
isotropic coordinates with metric \cite{kk1,kk2}
\begin{equation}
ds^2=
  - f dt^2 +  \frac{m}{f} d r^2 + \frac{m r^2}{f} d \theta^2 
           +  \frac{l r^2 \sin^2 \theta}{f} d\varphi^2
\ , \label{metric} \end{equation}
and $f$, $m$ and $l$ are only functions of $r$ and $\theta$.
For the gauge field a purely magnetic ansatz  
($A_0=0$) is used \cite{kk1,kk2}
\begin{eqnarray}
A_\mu dx^\mu =
\frac{1}{2er} \left[ \tau^n_\varphi 
 \left( H_1 dr + \left(1-H_2\right) r d\theta \right) \right.
\   \nonumber \\
 \left.
 -n \left( \tau^n_r H_3 + \tau^n_\theta \left(1-H_4\right) \right)
  r \sin \theta d\varphi \ \right]
\ . \label{gf} \end{eqnarray}
Here the symbols $\tau^n_r$, $\tau^n_\theta$ and $\tau^n_\varphi$
denote the dot products of the cartesian vector
of Pauli matrices, $\vec \tau = ( \tau_x, \tau_y, \tau_z) $,
with the spatial unit vectors
\begin{eqnarray}
\vec e_r^{\, n}      &=& 
(\sin \theta \cos n \varphi, \sin \theta \sin n \varphi, \cos \theta)
\ , \nonumber \\
\vec e_\theta^{\, n} &=& 
(\cos \theta \cos n \varphi, \cos \theta \sin n \varphi,-\sin \theta)
\ , \nonumber \\
\vec e_\varphi^{\, n}   &=& (-\sin n \varphi, \cos n \varphi,0) 
\ , \label{rtp} \end{eqnarray}
respectively.
Since the fields wind $n$ times around, while the
azimuthal angle $\varphi$ covers the full trigonometric circle once,
we refer to the integer $n$ as the winding number of the solutions.
For $n=1$ and $H_1=H_3=0$, $H_2=H_4=w(r)$,
the spherically symmetric ansatz of ref.~\cite{su2}
is recovered.

To fix the residual gauge degree of freedom \cite{kk1,kk2}
we choose the gauge condition 
\begin{equation}
r\partial_r H_1-\partial_\theta H_2 =0 \ .
\label{gc} \end{equation}

We consider static axially symmetric solutions
which are asymptotically flat, have a finite mass,
and are regular at the origin or possess a regular event horizon.
The boundary conditions at infinity and along the axes
are the same for the regular and black hole solutions.
At infinity ($r=\infty$) the boundary conditions are \cite{kk1,kk2}
\begin{equation}
f=m=l=1 \ , \ \ \ 
H_2=H_4=\pm 1, \ \ \ H_1=H_3=0
\ , \end{equation}
implying that the solutions are magnetically neutral.

Along the $\theta=0$-axis and the $\theta=\pi/2$-axis $H_1=H_3=0$; 
for all other functions the derivatives 
with respect to $\theta$ vanish on these axes \cite{kk1,kk2}.

For the regular solutions the boundary conditions at the origin 
($r=0$) are
\begin{eqnarray}
&&\partial_r f= \partial_r m= \partial_r l=0 
\ , \nonumber \\
&&H_2=H_4=1 \ , \ \ \ H_1=H_3=0
\ . \label{bc3} \end{eqnarray}

The regular horizon of the static black hole solutions 
resides at $r_{\rm H}$, where $g_{tt}=-f=0$ \cite{kk1,kk2}. 
The equations of motion impose the boundary conditions
at the horizon
\begin{eqnarray}
&&  f=m=l=0 
\ , \nonumber \\
&&   \partial_\theta H_1 + r \partial_r H_2 = 0 \ , \ \ \
 r \partial_r H_3-H_1 H_4=0
\ , \nonumber \\
&&   r \partial_r H_4+H_1( H_3 + {\rm ctg} \theta) =0 
\ . \label{bc4} \end{eqnarray}
For black hole solutions, the gauge condition (\ref{gc})
still allows for non-trivial gauge transformations.
The gauge is fixed by the condition \cite{kk1,kk2}
\begin{equation}
 \partial_\theta H_1 = 0 \ .
\end{equation}

\section{\bf Black Hole Properties}

Let us now discuss the properties of the non-abelian black holes.
Since they are static and carry no electromagnetic charge,
their only global charge is the mass.

The mass $M$ of the regular and black hole solutions
can be obtained directly from
the total energy-momentum ``tensor'' $\tau^{\mu\nu}$
of matter and gravitation,
$M=\int \tau^{00} d^3r$ \cite{wein}.

Changing to dimensionless coordinates, $x=(e/\sqrt{4\pi G}) r$,
the dimensionless mass $\mu =(e/\sqrt{4\pi G}) G M$
of the solutions is given by \cite{kk1,kk2}
\begin{equation}
\mu = \frac{1}{2} x^2 \partial_x f |_\infty
\ . \label{mass} \end{equation}

Introducing the dimensionless area parameter $x_\Delta$
of a black hole with horizon area $A$,
\begin{equation}
x_\Delta = \sqrt{A/4\pi} \ ,
\label{IHx} \end{equation}
we present the mass of the static black hole solutions, $\mu_{\rm bh}$,
with winding number $n$ and node number $k$
as a function of the area parameter in Fig.~1,
for the spherically symmetric solutions with $(n=1,k=1)$, $(n=1,k=2)$, 
and the axially symmetric solutions with $(n=2,k=1)$.
With increasing horizon size, the mass of the hairy black hole solutions
approaches the mass of the
Schwarzschild solutions.
\begin{figure}[h!]
\begin{center}
\epsfysize=6cm
\mbox{\epsffile{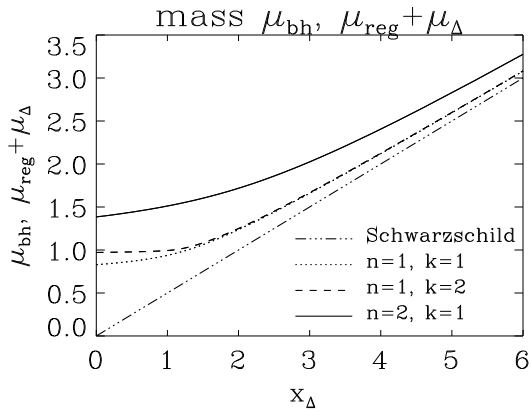}}                     
\caption{
 The dependence of the black hole mass $\mu_{\rm bh}$ 
 on the area parameter $x_{\Delta}$ 
 is shown for the hairy black hole solutions with
 $(n=1,k=1)$, $(n=1,k=2)$, $(n=2,k=1)$.
 For these solutions
 the sum $\mu_{\rm reg} + \mu_\Delta$ is superimposed (and cannot be distinguished
 graphically).
 Also shown is the mass of the Schwarzschild solutions.
}
\end{center}
\end{figure}                          
(Clearly, the non-abelian black holes are not characterized by their mass alone, but
need further specifications, such as e.g.~their winding and node numbers.)

Let us now address the horizon properties of the non-abelian black holes.

The zeroth law of black hole mechanics also holds 
for the static axially symmetric black hole solutions \cite{kk2}.
The surface gravity $\kappa_{\rm }$ \cite{ewein},
\begin{equation}
\kappa^2_{\rm }=-(1/4)g^{tt}g^{ij}(\partial_i g_{tt})(\partial_j g_{tt})
\ , \label{sg} \end{equation}
is constant on the horizon.

With $\partial^2_x f|_{x_{\rm H}} =2 f_2$ and
$\partial^2_x m|_{x_{\rm H}} =2 m_2$ 
we obtain for $\kappa$
\begin{equation}
\kappa=\frac{f_2}{\sqrt{m_2} } 
\ . \label{temp} \end{equation}

The isolated horizon framework \cite{iso} 
leads to an intriguing relation between the ADM mass $\mu_{\rm bh}$
of a black hole with area parameter $x_\Delta$,
and the ADM mass $\mu_{\rm reg}$
of the corresponding regular solution \cite{iso1},
\begin{equation}
\mu_{\rm bh} = \mu_{\rm reg} + \mu_\Delta \ , 
\label{IHmu} \end{equation}
where the horizon mass $\mu_\Delta$ is defined by
\begin{equation}
\mu_\Delta = \int_0^{x_\Delta} \kappa(x'_\Delta)x'_\Delta d x'_\Delta \ .
\label{IHmuD}
\end{equation}
This mass formula (\ref{IHmu}) has been shown previously
to hold for the numerically obtained
static spherically symmetric SU(2) solutions \cite{iso1}.
Here we show, that it also holds for the static axially symmetric solutions.
We demonstrate this relation in Fig.~1, where the sum
of the mass of the regular solution and the horizon mass,
$\mu_{\rm reg} + \mu_\Delta$, is superimposed on the
black holes mass, $\mu_{\rm bh}$,
for the EYM solutions with $(n=1,k=1)$, $(n=1,k=2)$, and $(n=2,k=1)$.
The curves agree within the numerical accuracy.

For distorted horizons, relation (\ref{IHmu}) has been verified previously
in Einstein-Yang-Mills-Higgs (EYMH) theory \cite{map3,hkk},
where it generalizes to include
bifurcating branches of non-abelian black hole solutions \cite{map3,iso2}.

The isolated horizon formalism also leads to a heuristic
physical model of hairy black holes,
which suggests to interpret a hairy black hole as a bound
state of a regular solution and a Schwarzschild black hole \cite{iso2}.
Rewriting relation (\ref{IHmu})
for the mass $\mu_{\rm bh}$ of the black hole solution 
with horizon radius $x_\Delta$ 
by introducing the mass $\mu_{\rm S} = x_\Delta/2$ 
of the Schwarzschild black hole with horizon radius $x_\Delta$,
\begin{eqnarray}
\mu_{\rm bh} 
 = \mu_{\rm reg} + \mu_{\rm S} + \left( \mu_\Delta - \mu_{\rm S} \right)
\ , \label{IHbs} \end{eqnarray}
yields for the binding energy $\mu_{\rm bind}$ of the system,
\begin{equation}
\mu_{\rm bind}= \mu_\Delta-\mu_{\rm S} \ .
\label{IHbind} \end{equation}

In Fig.~2 we show the binding energy $\mu_{\rm bind}$
for the spherically symmetric solutions with $(n=1,k=1)$, $(n=1,k=2)$, 
and the axially symmetric solutions with $(n=2,k=1)$.
When the binding energy tends to the negative value of the mass of the
regular solution for $x_\Delta \rightarrow \infty$, this confirms
the mass relation for the regular solution \cite{iso2},
\begin{equation}
\mu_{\rm reg} = \frac{1}{2} \int_0^{\infty} 
 \left( 1 - 2 \kappa(x'_\Delta)x'_\Delta \right) d x'_\Delta 
\ . \label{IHmuD2}
\end{equation}

\begin{figure}[h]
\begin{center}
\epsfysize=6cm
\mbox{\epsffile{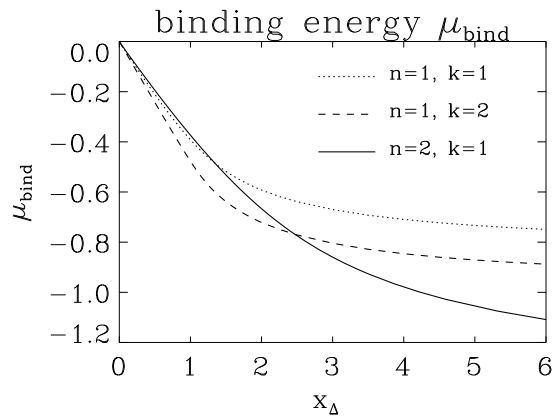}}                      
\caption{
 The dependence of the binding energy $\mu_{\rm bind}$
 on the area parameter $x_{\Delta}$ 
 is shown for the hairy black hole solutions with
 $(n=1,k=1)$, $(n=1,k=2)$, $(n=2,k=1)$.
}
\end{center}
\end{figure}                          

Let us now turn to the ``quasi-local uniqueness conjecture''
for black holes, put forward in ref.~\cite{iso1},
which states that static black holes are uniquely determined
by their horizon area and their horizon electric and magnetic charges.
Here the non-abelian electric and magnetic charge of the horizon \cite{iso,iso1}
appear as the crucial quantities.

The non-abelian magnetic charge of the horizon 
is defined via the surface integral over the horizon
\cite{iso,iso1}
\begin{equation}
P^{\rm YM}_\Delta = \frac{1}{4\pi} 
\oint \sqrt{\sum_i{\left(F^i_{\theta\varphi}\right)^2}} d\theta d\varphi 
\ , \label{Pdelta}
\end{equation}
and the non-abelian horizon electric charge is defined analogously
with the dual field strength tensor
\cite{iso,iso1}.

For the hairy black hole solutions in SU(2) EYM theory, the
horizon electric charge is identically zero. Only the horizon magnetic charge
is non-trivial and needs to be considered.

For the spherically symmetric solutions, the 
``quasi-local uniqueness conjecture'' holds \cite{iso1}.
For a given value of the area parameter, 
the horizon magnetic charge increases with increasing node number $k$,
reaching the constant value of the RN solution with unit charge
(for horizon $x_\Delta \ge 1$)
in the limit $k \rightarrow \infty$.

However, the conjecture should also hold
when the axially symmetric black hole solutions with
winding number $n>1$ are taken into account \cite{iso1}.
To address this crucial point, we have reanalyzed the
set of hairy black hole solutions with winding number $n=2$ and
node number $k=1$.
We exhibit the horizon magnetic charge $P_\Delta$ of these
non-abelian black hole solutions 
as a function of the area parameter $x_\Delta$ in Fig.~3,
together with the horizon magnetic charge of the
spherically symmetric solutions with one and two nodes.

\begin{figure}
\begin{center}
\epsfysize=6cm
\mbox{\epsffile{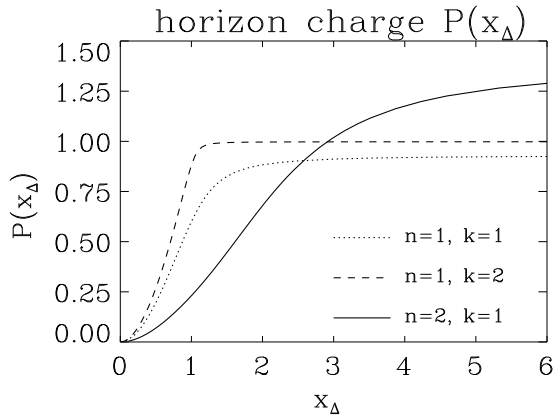}}                     
\caption{
 The dependence of the non-abelian horizon magnetic charge $P(x_\Delta)$
 on the area parameter $x_{\Delta}$ 
 is shown for the hairy black hole solutions with
 $(n=1,k=1)$, $(n=1,k=2)$, $(n=2,k=1)$.
}
\end{center}
\end{figure}                          

Whereas the curves of the spherically symmetric solutions do not intersect,
the curve of the axially symmetric $(n=2,k=1)$ solution
intersects the curves of both spherically symmetric solutions.
In fact, it intersects the curves of all the spherically symmetric solutions,
since they increase monotonically with $k$, but have $P_\Delta < 1$.
Furthermore it also intersects the RN curve, $P_\Delta=1$
(for $x_\Delta \ge 1$).

Thus when black hole solutions with a distorted horizon are included,
the ``quasi-local uniqueness conjecture'' holds no longer.
Therefore the need for a modified uniqueness conjecture persists.
The set of two integers $(n,k)$, used to characterize the hairy black holes
with a given mass or horizon area, apparently cannot be replaced
by the single quantity of the horizon magnetic charge.
Another horizon property would have to be included 
in the uniqueness conjecture. 
The deformation of the horizon could serve to 
distinguish black holes solutions with equal horizon area
and horizon magnetic charge \cite{foot}.

We will revisit the uniqueness conjecture in connection with rotating EYM black holes
\cite{rot}.


\begin{thebibliography}{000}
\bibitem{foot}
 Deformation parameters
 could be the length of the equatorial circumference of the horizon $L_e$,
 the length of the polar circumference $L_p$, or their ratio $L_e/L_p$
 \cite{kk2}.
\end{thebibliography}
\end{document}